\documentstyle[12pt,fleqn]{article}
\def\beq{\begin{equation}}
\def\eeq{\end{equation}}
\def\6{\langle}
\def\9{\rangle}

\begin{document}
\noindent Proceedings of the Adriatico Research Conference ``Quantum
Interferometry III''\\ Trieste, March 1999 (to be published in {\sl
Fortschritte der Physik\/}).\\[10mm]

{\Large{\bf Bayesian Analysis of Bell Inequalities}}\bigskip

{\sc Asher Peres} \bigskip

Department of Physics, 

Technion---Israel Institute of Technology, 

32\,000 Haifa, Israel

Electronic address: peres@photon.technion.ac.il\\[15mm]

{\bf Abstract}\bigskip

\noindent Statistical tests are needed to determine experimentally
whether a hypothetical theory based on local realism can be an
acceptable alternative to quantum mechanics. It is impossible to
rule out local realism by a single test, as often claimed erroneously.
The ``strength'' of a particular Bell inequality is measured by the
number of trials that are needed to invalidate local realism at a given
confidence level. Various versions of Bell's inequality are compared
from this point of view. It is shown that Mermin's inequality for
Greenberger-Horne-Zeilinger states requires fewer tests than the
Clauser-Horne-Shimony-Holt inequality or than its chained variants
applied to a singlet state, and also than Hardy's proof of
nonlocality.\\[10mm]

\parindent 5mm \mathindent 5mm
\noindent{\bf I.\hskip15mm Formulation of the problem}\bigskip

\noindent Bell inequalities are upper bounds on the correlations of
results of distant measurements. These inequalities are obeyed by any
local realistic theory, namely a theory that uses local variables with
objective values. Since Bell's original discovery~[1], many inequalities
of that type have been published, with various claims of superiority.
The purpose of this article is to compare their relative strengths for
various quantum states.

In actual experimental tests, there are no infinite ensembles for
accurate measurements of mean values.  Experimental physicists perform a
finite number of tests, and then they state that their results violate
the inequality at some confidence level. The problem I wish to discuss
here is of a different nature. I am a theorist and I trust that quantum
mechanics gives a reliable description of nature. However, I have a
friend who is a local realist. We have only a finite number of trials at
our disposal. How many tests are needed to make my realist friend feel
uncomfortable? 

The problem is not whether the validity of a Bell inequality can be
salvaged by invoking clever loopholes, as some local realists try to
trick us into, but whether there can be any local realistic theory that
reproduces the experimental results. When these results are analyzed we
have to take into account detector inefficiencies, and this should be
done honestly in the same way when our analysis is based on quantum
theory or on a local realistic theory. To simplify the discussion, I
shall assume that there are ideal detectors, and that the rate at which
particles are produced by the apparatus is perfectly known. The
disagreement is only on the choice of the correct theory. 

Consider a yes-no test. Quantum mechanics (QM) predicts that the
probability of the ``yes'' result is $q$, and an alternative local
realistic (LR) theory predicts a probability $r$. An experimental test
is performed $n$ times and yields $m$ ``yes'' results. What can we infer
about the likelihood of the two theories? The answer is given by Bayes's
theorem~[2]. Denote by $p'_q$ and $p'_r$ the prior probabilities that we
assign to the validity of the two theories. These are subjective
probabilities, expressing our personal beliefs. For example, if my
friend is willing to bet 100 to~1 that LR is correct and QM is wrong,
then $p'_r/p'_q=100$. The question is how many experimental tests are
needed to change my friend's opinion to $p''_r/p''_q=0.01$ say, before
he is driven to bankruptcy.

It follows from Bayes's theorem that

\beq {p''_r\over p''_q}={p'_r\over p'_q}\;{E_r\over E_q}, \eeq
where $E_r$ and $E_q$ are the probabilities of the experimentally found
result (namely $m$ successes in $n$ trials), according to the two
theories. These are, by the binomial theorem,

\beq E_r=[n!/m!(n-m)!]\;r^m\,(1-r)^{n-m},\eeq
\beq E_q=[n!/m!(n-m)!]\;q^m\,(1-q)^{n-m},\eeq
whence

\beq E_q/E_r=(q/r)^m\;[(1-q)/(1-r)]^{n-m}. \label{conf} \eeq
I shall call the ratio 

\beq D=E_q/E_r \eeq
the {\it confidence depressing factor\/} for hypothesis LR with respect
to hypothesis QM.

\bigskip\noindent{\bf II.\hskip15mm Greenberger-Horne-Zeilinger
state}\bigskip

\noindent As a first example, consider the Greenberger-Horne-Zeilinger
(GHZ) state [3,~4] for a tripartite system, namely
$(|000\9-|111\9)/\sqrt{2}$, where 0 and 1 denote two orthogonal states
of each subsystem. This state is experimentally difficult to produce but
its theoretical analysis is quite simple. Three distant observers
examine the three subsystems. The first observer has a choice of two
tests.  The first test can give two different results, that we label
$a=\pm1$, and likewise the other test yields $a'=\pm1$. Symbols $b, b',
c$ and $c'$ are similarly defined for the two other observers. Any
possible values of their results satisfy

\beq a'bc+ab'c+abc'-a'b'c'\equiv\pm2, \label{pm2}\eeq
whence it follows that

\beq -2\leq\6a'bc+ab'c+abc'-a'b'c'\9\leq2.\label{mermin}\eeq
This is Mermin's inequality~[5]. 

Quantum mechanics happens to make a very simple prediction for the GHZ
state: there are well chosen tests that give with certainty

\beq a'bc=ab'c=abc'=-a'b'c'=1. \eeq
Naturally, performing any such test can verify the value 1 for only one
of these products, since each product corresponds to a different
experimental setup. Yet, if we take all these results together they
manifestly conflict with Eq.~(\ref{pm2}), and many authors~[6] have
stated that a single experiment is sufficient to invalidate local
realism. This is sheer nonsense: a single experiment can only verify one
occurrence of one of terms in (\theequation).

Let us return to our realist friend. He believes that, in each
experimental run, each term in Eq.~(\theequation) has a definite value
even if that term is not actually measured in that run. Let us
therefore ask him to propose just a rule giving the {\it average\/}
values of the products $a'bc$, etc., that appear in Eq.~(\ref{mermin}).
How many tests are needed for depressing his confidence in that rule by
a factor $10^4$, say?

The most successful LR theory, namely the one that gives the least
depressing factor, is to assume that 

\beq \6a'bc\9=\6ab'c\9=\6abc'\9=\6-a'b'c'\9=0.5. \eeq
This obviously attains the right hand side of Mermin's inequality
(\ref{mermin}). The LR prediction thus is that if we measure $a'bc$, we
shall find the result 1 (i.e., ``yes'') in 75\% of cases, and the
opposite result in 25\%; and likewise for the other tests. We thus have,
with the notations introduced above, $q=1$ and $r=0.75$. For $n$ tests,
with ideal detectors, we have $m=n$ (I am assuming here that quantum
theory is correct), and the depressing factor in Eq.~(\ref{conf}) is
$0.75^n$. For example, 32 tests give $D\simeq10^4$, as required.

\bigskip\noindent{\bf III.\hskip15mm The singlet state}\bigskip

\noindent The second example involves just two correlated quantum
systems far away from each other. An observer, located near one of the
systems, has a choice of several yes-no tests, labelled $A_1$, $A_3$,
$A_5$, etc. Likewise, another observer, near the second system, has a
choice of several yes-no tests, $B_2$, $B_4$, $B_6$ \ldots\ Let
$p(A_iB_j)$ denote the probability that tests $A_i$ and $B_j$ give the
same result (both ``yes'' or both ``no''). It was shown long ago by
Clauser, Horne, Shimony, and Holt (CHSH)~[7] that local realism implies  

\beq p(A_1B_2)+p(B_2A_3)+p(A_3B_4)\ge p(A_1B_4). \eeq 
(In the original paper [7], this equation was written in terms of {\it
correlations\/}, namely $2p-1$, but it is much simpler to use
probabilities, as here.)  More generally, Braunstein and Caves [8]
derived chained Bell inequalities that can be written  

\beq p(A_1B_2)+p(B_2A_3)+\cdots+p(A_{2k-1}B_{2k})\ge p(A_1B_{2k}). \eeq
There are $(k!)^2$ independent inequalities of that type, obtainable by
relabelling the various $A_i$ and $B_j$. Local realism guarantees that
all these inequalities are satisfied.  

Consider a pair of spin-$1\over2$ particles in the singlet state
(similar results hold for maximally entangled pairs of polarized
photons, except that all the angles mentioned below should be halved).
Each observer can measure a spin component along one of $k$ possible
directions, as illustrated in Fig.~1, where the angle between
consecutive directions is $\theta=\pi/2k$. Quantum theory predicts that
each one of the probabilities on the left hand side of
Eq.~(\theequation) is $q=(1-\cos\theta)/2$, and the probability on the
right hand side is $1-q$. These predictions manifestly violate
Eq.~(\theequation).  

What could be the predictions of an alternative theory, based on local
realism? These predictions have to satisfy Eq.~(\theequation). The
closest they can approach quantum theory is when equality holds in the
latter equation. Moreover, it is reasonable to assume that all the terms
on the left hand side are equal (this follows from rotational symmetry,
and it can also be shown that any deviation from this symmetry would
only increase the depression factor $D$). Let $r$ be the common value of
all these terms. Then the right hand side of Eq.~(\theequation) has to
be $1-r$ (again, because of rotational symmetry and because the spin
projection along $\beta_{2k}$ is opposite to that along $\beta=0$). It
follows that in a local realistic theory which mimics as closely as
possible quantum mechanics and saturates the inequality (\theequation),
we have $(2k-1)r=1-r$, whence $r=1/2k=\theta/\pi$, where $\theta$ is the
angle between consecutive rays. This is indeed the result obtainable
from a crude semi-classical model, where a spinless system splits into
two fragments with opposite angular momenta~[9]. Quantum theory, on the
other hand, predicts for the same angle $\theta$ a probability
$q=(1-\cos\theta)/2$ that both observers obtain the same result. 

We thus have now definite predictions, from quantum theory and from an
alternative local realistic theory. To distinguish experimentally
between these two claims, we test $n$ pairs of particles prepared in the
singlet state. Let $m$ be the number of ``yes'' answers. If $m\simeq
qn$ (that is, if quantum theory is experimentally correct), it follows
from Eq.~(\ref{conf}) that

\beq D=\left[\left({q\over r}\right)^q\;
  \left({1-q\over 1-r}\right)^{1-q}\right]^n. \label{D} \eeq
For example, if we wish to have $D\simeq10^4$ as before, we obtain
$n\simeq 287$ for $k=2$ (the case that was investigated by CHSH), and
$n\simeq 200$ for the more efficient configuration of Braunstein and
Caves with $k=4$ (higher values of $k$ would require a higher number
of tests for giving the same depression factor $D$). 

\bigskip\noindent{\bf IV.\hskip15mm Hardy's proof of
nonlocality}\bigskip

Finally, let us examine Hardy's proof of nonlocality ``without
inequalities'' [10,~11], which was called by Mermin ``the best version
of Bell's inequality''~[12]. It will be shown that this version is not
stronger than the preceding ones. Stripped of all its technical details,
Hardy's paradox can be formulated as follows. There are four alternative
setups as in the CHSH case, but each setup requires only one detector.
The first observer has a choice of using detectors $A$ or $A'$, the
second observer may use $B$ or $B'$. Detector coincidences will be
labelled $C_j$, with $j=1,\ldots,4$. Explicitly,

\beq C_1=A\wedge B, \qquad\qquad C_2=A\wedge B', \qquad\qquad
  C_3=A'\wedge B, \eeq
and $C_4$ means that in the fourth setup neither $A'$ nor $B'$ is
excited. Other types of coincidences are not relevant in the following
discussion.

Local realism implies that the probabilities $p(C_j)$ satisfy the
Clauser-Horne (CH) inequality~[13], 

\beq p(C_1)\leq p(C_2)+p(C_3)+p(C_4). \label{CH}\eeq
On the other hand, quantum mechanics predicts that, for well chosen
states and tests, these probabilities are $p(C_2)=p(C_3)=p(C_4)=0$ and 

\beq  p(C_1)\equiv q=[(\sqrt{5}-1/2)]^5=0.09017, \eeq
so that the CH inequality is violated.

As in the preceding cases, let our LR friend propose a new set of
probabilities $r_j$ that satisfy the CH inequality. For example, a
simple possibility is to postulate that all the $r_j$ vanish (this
assumption is implicit in Hardy's proof). Then LR and QM agree for
setups 2, 3, and 4, and we only have to test experimentally setup~1.
According to QM, the probability of finding $n$ consecutive ``no''
results (in agreement with the LR prediction) is $(1-q)^n$. This is less
than 50\% after only 8 trials. The hypothesis that all the $r_j$ vanish
is obviously untenable, and this is why Hardy's proof is usually
considered as quite strong. 

However, there is a more sophisticated way to defend local realism. Let
us assume that $r_2=r_3=r_4=r_1/3$, so that the CH inequality~(\ref{CH})
is saturated, and let us optimize the value of $r_1\equiv r$. There are
now two types of experimental tests. Those with setup~1 lead to a value
of $D$ given by Eq.~(\ref{D}). On the other hand, setups 2, 3, and 4
have $q=0$ and then Eq.~(\ref{D}) gives, with $r$ replaced by $r/3$, the
result $D=(1-r)^{-n}$. To invalidate local realism, we shall obviously
choose the setup that minimizes $n$, the number of required tests.
Therefore the best that a LR theorist can do is to choose $r$ so as to
equate these two values of $D$. A straigthforward calculation then gives
$r=0.03358$ and $n\simeq270$. 

\bigskip\noindent{\bf Acknowledgments}\bigskip

\noindent I am grateful to Dagmar Bru\ss\ for pointing out ambiguities
in an earlier version of this article, and to Chris Fuchs for helpful
comments. This work was supported by the Gerard Swope Fund and the Fund
for Encouragement of Research.\clearpage

\frenchspacing \begin{enumerate}
\item {\sc J. S. Bell}, Physics {\bf1}, 195 (1964).
\item {\sc L. J. Savage}, {\it The Foundations of Statistics\/}, (Dover,
New York, 1972).
\item {\sc D. M. Greenberger, M. A. Horne,} and {\sc A. Zeilinger}, in
{\it Bell's Theorem, Quantum Theory and Conceptions of the Universe\/},
ed.\ by {\sc M. Kafatos} (Kluwer, Dordrecht, 1989).
\item {\sc N. D. Mermin}, Physics Today {\bf43} (6), 9 (1990); Am. J.
Phys. {\bf58}, 731 (1990).
\item {\sc N. D. Mermin}, Phys. Rev. Letters {\bf65}, 1838 (1990).
\item The list of authors is too long to be given explicitly, and it
would be unfair to give only a partial list.
\item {\sc J. F. Clauser, M. A. Horne, A. Shimony}, and {\sc R. A. Holt},
Phys. Rev. Letters {\bf23}, 880 (1969).
\item {\sc S. L. Braunstein} and {\sc C. M. Caves}, Ann. Phys. (NY) {\bf
202}, 22 (1990).
\item {\sc A. Peres}, {\it Quantum Theory: Concepts and Methods\/}
(Kluwer, Dordrecht, 1993) p.~161.
\item {\sc L. Hardy}, Phys. Rev. Letters {\bf71}, 1665 (1993).
\item {\sc S. Goldstein}, Phys. Rev. Letters {\bf72}, 1951 (1994).
\item {\sc N. D. Mermin}, in ``Fundamental Problems in Quantum Theory''
ed. by {\sc D. M. Greenberger} and {\sc A. Zeilinger}, Ann. New York
Acad.~Sci. {\bf755}, 617 (1995).
\item {\sc J. F. Clauser} and {\sc M. A. Horne}, Phys. Rev. D {\bf10},
526 (1974).

\end{enumerate}

\vfill\noindent Fig. 1. \ The Braunstein-Caves configuration for chained
Bell inequalities: there are $k$ alternative directions along which each
observer can measure a spin projection. 
\end{document}